\documentclass[twocolumn,prb,showpacs]{revtex4}

\usepackage{graphicx}
\usepackage{rotating}
\usepackage{amsmath}
\usepackage{amsfonts}
\usepackage{amssymb}
\usepackage{enumerate}
\usepackage{longtable}
\usepackage{dcolumn}
\usepackage{bm}

\begin{document}

\newcommand{\be}{\begin{equation}}
\newcommand{\ee}{\end{equation}}
\newcommand{\bn}{\begin{eqnarray}}
\newcommand{\en}{\end{eqnarray}}

\title{Anomalous Magnetic Susceptibility in Iron Pnictides: Paramagnetic Phase}

\author{M.S. Laad and L. Craco}
\affiliation{Max-Planck-Institut f\"ur Physik komplexer Systeme,
01187 Dresden, Germany}

\date{\rm\today}

\begin{abstract}
Observation of an anomalous temperature dependence of the spin susceptibility, 
along with a spin gap in NMR, in the quantum paramagnetic 
{\it normal} state of Iron Pnictides is a signature of an unusual metallic
state.  We argue 
that {\it both} these anomalous features are associated with a wide 
fluctuational regime dominated by dynamical, short-ranged and frustrated 
spin correlations in a strongly correlated metal. Using LDA+DMFT, we show 
that {\it both} these features can be quantitatively undertstood in the 
doped Fe-pnictides. We argue that such spin correlations naturally arise 
in a {\it Mottness} scenario, where an {\it effective}, dualistic description 
involves coexisting renormalized quasiparticles and effectively localized 
moments, arising from the same set of $d$-bands. 
\end{abstract}
   

\maketitle

The precise mechanism of unconventional superconductivity (U-SC) in the
recently discovered Iron Pnictides (FePn) is presently a hotly debated
issue.~\cite{[1]} U-SC arises on the border of a geometric 
frustration (GF)-induced ``stripe''-spin-density-wave (SSDW).~\cite{[2]} 
Short-ranged, frustrated spin correlations, expected to 
survive doping-induced destruction of SSDW order,~\cite{[3]} are 
then expected to play an important role in the normal and SC phases of FePn. 

Study of magnetic fluctuations in an U-SC can help unearth the symmetry
of the SC order parameter, as shown by detailed studies for cuprates~\cite{[4]}.
In FePn, extant experiments already show anomalous behavior. Specifically, 
NMR studies reveal normal state pseudogap behavior~\cite{[5]} and U-SC, as 
seen from opening up of a spin gap at $T^{*}=150-200$~K, and 
$T_{1}^{-1}\simeq T^{n}$ with $n=2.2-2.5$ for $T<<T_{c}$. However, other 
probes reveal anisotropic, albeit fully gapped, structure of the in-plane 
gap function.  While this is the expected response of a GF spin system in 
its quantum disordered phase,~\cite{[2]} FePn are {\it metals}, albeit 
presumably close to a Mott-Hubbard instability.~\cite{[2]}  
Are there other signatures of anomalous magnetic correlations in FePn?  
Extant data reply in the affirmative: the {\it uniform} spin
susceptibility shows a {\it linear}-in-$T$ dependence for $T>T_{SDW}(x)$ up to
$800$~K for {\it both} the 1111- and 122-FePn.~\cite{[6],[kling]} This 
anomalous behavior persists upon destruction of the SSDW, even going over to 
a $\chi(T)\simeq T^{1+n}$ form at lower $T$, with $n\le 1$.  Actually,
just above $T_{c}$ for the doped La-based 1111-FePn with $x>0.05$, the spin
susceptibility obeys the $\chi(T)=\chi_{0}+AT^{1+n}$ law, and the 
linear-in-$T$ dependence is recovered smoothly only at higher $T$.
This observation is strong circumstantial evidence for relevance of GF, 
since the $\chi(T)\simeq T$ law is also seen above $T_{N}$ in another GF 
system, $Na_{0.5}CoO_{2}$, which is also a poor metal like doped 
FePn.~\cite{[7]}  However, $Cr$ and its alloys, widely believed to be described in a more itinerant framework, also show the linear-in-$T$ susceptibility.  The distinguishing feature of the FePn with $x\simeq 0.1$ is, however, that they 
exhibit the spin gap in NMR in the quantum {\it disordered} phase.  This observation puts them into the ``strong coupling'' category, since the spin gap in the
paramagnetic phase is beyond reach of an itinerant description.
A consistent theory of magnetic fluctuations in FePn 
{\it must} then reconcile the spin-gap in NMR with $\chi(T)\simeq T^{1+n}$, 
with $0\le n\le 1$ in a single picture.   

Earlier theoretical studies have focussed on both itinerant weak- 
and strong coupling routes.~\cite{[8],[6]} In the itinerant view, a 
$\chi(T)\simeq T$ {\it requires} 
nesting of the electron- and hole-like Fermi sheets. 
This involves fine tuning, since no such nesting 
can be invoked for doped 1111-FePn, even as the $\chi(T)\simeq T$ behavior 
persists.  
The second view, based on 
(presumably) Kondo-like coupling of itinerant carriers to effectively 
{\it localized} spins in a $J_{1}-J_{2}$-Heisenberg model, does not 
require fine tuning, but is strictly valid only in the localized regime.~\cite{[2]} 
Experimental evidence, however, places the FePn in the intermediate-to-strong 
correlation regime,~\cite{[2]} where the {\it dualistic} aspect of 
self-consistently correlated $d$ electrons should manifest itself. In this 
{\it Mottness} scenario, i{\it both} the ``itinerant'' carriers and the 
``local'' moments arise from the same set of 
$d$-electrons, simultaneously giving renormalized LDA-like ``bands'' as 
seen in dHvA~\cite{[9]} {\it and} ``localized'' magnetic responses seen 
in NMR and INS studies.  We thus 
use LDA+DMFT, which is known to capture ``Mottness'' 
{\it exactly} at the mean-field level.~\cite{[10],[11]}
     
In a study of magnetic fluctuations, the central quantity of 
interest is the {\it dynamical} spin susceptibility,
$\chi({\bf q},\omega)=\sum_{a,b}\chi_{a,b}({\bf q},\omega)$, where $a,b$ 
are {\it all} $d$-orbital indices, and ${\bf q},\omega$ are the momentum and
energy transfers in INS.  Viewing FePn as strongly correlated
systems, we construct $\chi({\bf q},\omega)$ in terms of the {\it full}
LDA+DMFT propagators computed in earlier work.~\cite{[12],[13],[19]} Good
quantitative agreement between LDA+DMFT and key experiments in {\it both}, the
normal and U-SC states, has been shown there, supporting our
choice.  The prescription is to replace the band Green functions used
in weak-coupling RPA-like approaches~\cite{[13]} by their LDA+DMFT counterparts.
This ensures that the dynamical aspect of strong, local, multi-orbital (MO) 
correlations is included from the outset. 

For the incoherent ``normal'' metallic state in FePn, after replacing the 
{\it bare} $G_{aa}({\bf k},\omega)$ with
$G_{aa}({\bf k},\omega)\equiv G_{aa}^{LDA+DMFT}({\bf k},\omega)=
[\omega-\epsilon_{ka}-\Sigma_{a}(\omega)]^{-1}$,
 and introducing the spin operator
$S_{a,\mu}({\bf q})=\frac{1}{2}\sum_{\bf k}c_{a,\mu,\sigma}^{\dag}({\bf
k}+{\bf q}){\bf \sigma}_{a,\sigma,\sigma'}^{\mu}c_{a,\mu,\sigma'}({\bf k})$,
with $\mu=x,y,z$, the ``bare'' dynamical spin susceptibility reads

\bn
\nonumber
&&\chi_{0,a,b}^{\mu\nu}({\bf q},\omega)=
K_{\mu\nu}^{\sigma\sigma'} \int d\nu\int d\epsilon \sum_{\bf k,\omega'}
\rho_{aa\sigma}({\bf k}+{\bf q},\nu) \\ \nonumber
&& \rho_{bb\sigma'}({\bf k},\epsilon)\frac{n_{F}(\nu)-n_{F}(\epsilon)}{\omega+\nu-\epsilon+i\eta},
\en
where
$K_{\mu\nu}^{\sigma\sigma'}\equiv\frac{1}{2}{\sigma}_{a,\sigma\sigma'}^{\mu}\cdot{\sigma}_{b,\sigma\sigma'}^{\nu}$, the $\sigma$ are Pauli matrices, and
$n_{F}$ the Fermi function.  $\rho_{\alpha}({\bf k},\nu)=(-1/\pi)$Im$G_{\alpha\alpha}({\bf k},\nu)$ is the one-particle spectral function for orbital $\alpha$.
Including the ladder vertex in an infinite summation of ``ladder'' diagrams
using RPA, the renormalized magnetic susceptibility, $\chi_{a,b}({\bf q},\omega)=[\chi_{0,a,b}^{-1}({\bf q},\omega)-J({\bf q})]^{-1}$, where, following~\cite{[rmpg]}
$\chi_{0,a,b}({\bf q},\omega)$ is evaluated in DMFT.  The ``bare'' bubble
contribution is found as $\chi^{(0)}(0,\omega)\simeq \chi^{(0)}(\omega)=C\sum_{a,b}\int d\nu \rho_{a}(\nu)\rho_{b}(\omega+\nu)[f(\nu)-f(\omega+\nu)]$, and
$J({\bf q})=J_{1}$(cos($q_{x}a$)+cos($q_{y}a$))+$J_{2}$cos($q_{x}a$)cos($q_{y}a$),
with $J_{1}\simeq \frac{t_{ab}^{2}}{U'+J_{H}}$ and $J_{2}\simeq\frac{t_{ab}'^{2}}{U'+J_{H}}$ being the frustrated superexchange scales in
FePn.~\cite{[2]} Using $\chi_{0,a,b}({\bf q},\omega)$, the dynamical spin 
susceptibility, ${\chi''({\bf q},\omega)}$, can be now expressed in terms of 
the {\it full} DMFT propagators computed earlier for the incoherent ``normal'' 
state.~\cite{[12]} 

Using $\chi({\bf q},\omega)$ at low $T$, the uniform (${\bf q}=0$) spin 
susceptibility is now estimated as 
\be
\nonumber
\chi(T)\equiv\chi(0,T)=\int_{-\infty}^{+\infty}\frac{\chi''(0,\omega)d\omega}{1-e^{-\beta\omega}}.
\ee
Notice how frustration appears in the ``bare'' 
bubble comprised of fully renormalized DMFT propagators, which contain 
contributions from the (frustrated) hoppings 
(via DMFT propagators), as well as in $J({\bf q})$ via the RPA sum. 
In a companion work,~\cite{[13]} 
we have shown how this yields nice agreement with the NMR relaxation rate 
over the {\it full} $T$ range for FePn,~\cite{[13]} including spin gap 
formation {\it and} unconventional power-law form of $T_{1}^{-1}$ below 
$T_{c}$.  Here, we extend that study to investigate the spin susceptibility 
in detail.  We explicitly demonstrate that $\chi(T)\simeq T$ at high $T$, 
with a smooth crossover to a $\chi(T)\simeq T^{1+n}$ form 
at lower $T$ with $n$ smoothly increasing from zero to unity as $T$ is 
lowered, for the doped FePn which become superconducting at lower $T$.  
We show how {\it both} NMR and susceptibility data in the ``normal'' state
can be rationalized in terms of strong, frustrated spin correlations arising 
in a multi-band, correlated system on the verge of {\it Mottness}.    

Before presenting our results, we make a few relevant remarks.  In contrast 
to earlier studies,~\cite{[6],[8]} our model incorporates {\it all} five $d$
bands, and explicitly captures the anisotropic three-dimensional band structure
of the 1111-FePn.  This should become relevant at low $T$, and a {\it smooth} 
crossover to $D=2$ physics should occur with increasing $T$ above
$T_{SDW}(x)$.  In such a $D=2$ regime, one might associate the 
$\chi(T)\simeq T$ law with the fluctuational regime that occurs {\it below} 
the mean-field ($T_{MF}\simeq J_{2}>>T_{SDW},T_{SC}$) crossover.~\cite{[6]}  
The wide $T$ range over which these latter exist points to dominant effects of
non-local spatial spin correlations up to rather high $T$.  Combining this with
the GF nature of FePn {\it and} observation of the spin gap below $T^{*}$ in 
the 1111-FePn suggests that the spatial extent of these spin correlations 
should be of order a $Fe-Fe$ unit cell. These are precisely the correlations 
accessed by our formulation above, though our neglect of vertex corrections 
of the ``crossing diagram'' type is an approximation.  We note, however, 
that a similar approximation is shown to reconcile the ARPES lineshapes 
{\it and} the neutron scattering results in cuprates.~\cite{[14]}  This 
indicates that such vertex corrections are small, justifying our approximation.

\begin{figure}[thb]
\includegraphics[width=\columnwidth]{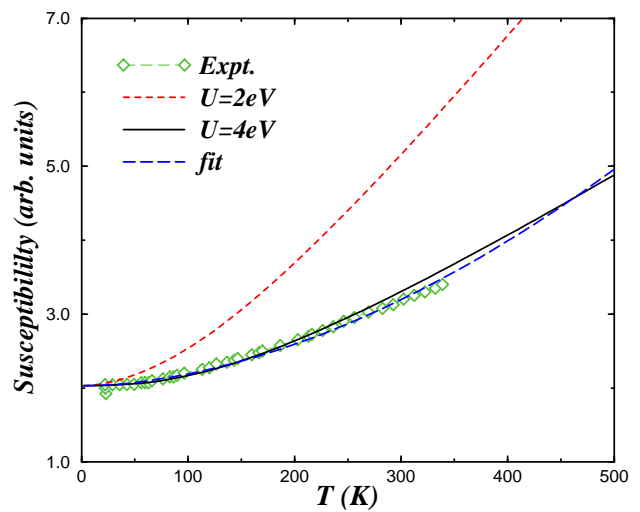}
\caption{(Color online) 
$T$-dependence of the uniform spin susceptibility, $\chi(T)$, for the $x=0.1$
La-based Fe-pnictide.  Notice the {\it smooth} evolution of the linear-in-$T$
law at ``high''-$T>200$~K to a $T^{1.8}$ law at lower $T$, in nice agreement
with experiment.~\cite{[kling]}  Also note that the $U$-dependence of the
susceptibility is strong evidence for a sizably correlated multi-orbital
scenario in FePn.  For clarity, both theoretical curves are shifted downward
to coincide with the experimental curve at $T_{c}$.}
\label{fig1}
\end{figure}

\begin{figure}[thb]
\includegraphics[width=\columnwidth]{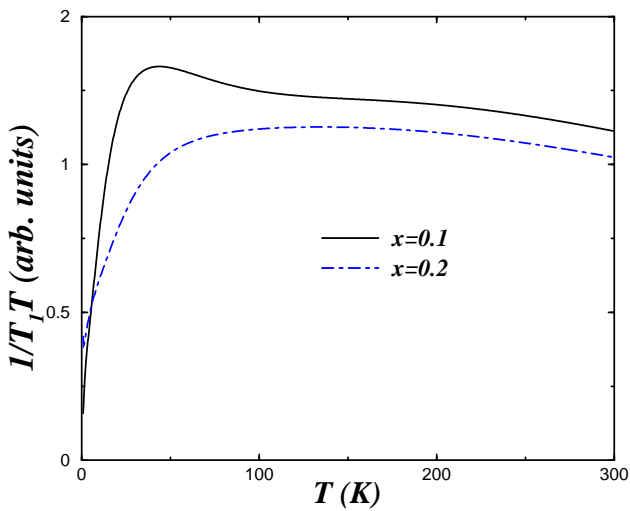}
\caption{(Color online) 
Low $T$ behavior of the NMR $[T_{1}T]^{-1}$ for $LaO_{1-x}FeAsF_{x}$ with
$x=0.1,0.2$, corresponding to doped samples with no SSDW order.  
The emergence of the spin gap around $T^{*}\simeq 150-200$~K
for the doped FePn is clearly resolved, in good agreement with
experiment.~\cite{[4]}}
\label{fig2}
\end{figure}

We now present our results.  In Fig.~\ref{fig1}, we show the $T$-dependent
uniform spin susceptibility, $\chi(T)$, for $LaO_{1-x}FeAsF_{x}$ at a 
representative electron doping, $x=0.1$, for two values of $U$ corresponding
to intermediate-to-strong ($U=4.0$~eV) and to weak ($U=2.0$~eV) coupling.  
Since we do not treat the SSDW instability here, we focus only on the 
regime $x>0.04$, where we study the $T$-dependent $\chi(T)$ up to lower 
$T$ in the ``normal'' state.  
For $x=0.1$, SSDW order is destroyed by doping.
However, in full agreement with experiment, $\chi(T)$ continues to show
the linear-in-$T$ dependence in the range $200$~K$<T<800$~K (only shown up to 
$350$~K), even exhibiting a smooth {\it increase} of curvature 
($\simeq T^{1.8}$) at even lower $T$ in the vicinity of $T_{c}$. In fact, 
for $x=0.1$, we have fitted $\chi(T)$ up to $500$~K with the
$2.03+0.42T^{1.8}$ law, quite distinct from the linear-in-$T$
behavior seen for small $x$.~\cite{[kling]})  Only at high $T$ is 
the linear-in-$T$ behavior recovered for $x\simeq 0.1$.  The $T$-dependence of 
our computed susceptibility is very close to this fit to experiment, 
as shown in Fig.~\ref{fig1}.  
Clearly, $\chi(T)$ with $U=2.0$~eV substantially deviates
from experiment, while a much better agreement is obtained with
$U=4.0$~eV, testifying to the importance of sizable multi-orbital electronic
corrrelations in FePn.  For sake of discussion, we show our earlier 
theoretical result for the NMR relaxation rate, $[T_{1}T]^{-1}$,~\cite{[13]} 
with $U=4.0$~eV in Fig.~\ref{fig2}.  The absence of a Korringa form 
($[T_{1}T]^{-1}$=const) at any $T$ is striking.  Around $T^{*}\simeq 150$~K,
clear, gradual opening up of a spin gap, again as seen in NMR work, is found.
The correlation between the spin-gap in NMR and $\chi(T)\simeq T$ is not 
easily obvious: $\chi(T)$ continues to vary {\it smoothly} even as 
$[T_{1}T]^{-1}$ shows the spin gap behavior.    
  
Such semi-quantitative agreement, even in details, with {\it both} NMR and
spin susceptibility experiments is
a very gratifying feature of our modelling.  Deeper insight into 
these anomalous responses, and of their connection to U-SC in doped FePn, is 
clearly desirable.  To this end, we start by noticing that $\chi(T)\simeq T$ is
found in the range $0<T<T_{MF}\simeq O(J)$ even for a classical, 
{\it unfrustrated} lattice in $D=2$.~\cite{[15]}  Since the $D=2$ Heisenberg 
model cannot have a finite $T_{N}$, it follows that this linear behavior must 
be associated with the fluctuational regime below the mean-field crossover, as
discussed by Zhang {\it et al.} in the localized limit of the $J_{1}-J_{2}$ 
model.~\cite{[6]} In a GF system like FePn, this regime involves short-ranged,
{\it dynamical} spin correlations above $T_{SDW}$.   
In ``bad metallic'' FePn, the ``Mottness'' philosophy allows us to describe 
the low-energy magnetic fluctuations by an {\it effective} $J_{1}-J_{2}$ 
model.~\cite{[2]}  In this model, there is a wide $T$ range,
$T_{SDW}<T<T_{s}$, where translational symmetry is spontaneously broken but 
spin-rotational SU$(2)$ invariance is not.~\cite{[2]}  Such a transition has 
interesting consequences:

(i) it naturally accounts for opening up of a spin gap, leading to consistency
with NMR data, and,

(ii) it breaks the symmetry between  ${\bf q}=(\pi,0)$ and $(0,\pi)$, already
above $T_{SDW}$. This breaks the lattice {\it rotational} symmetry 
(Ref.~\onlinecite{[3]}) and should lead to the tetragonal-orthorhombic 
(T-O) distortion,~\cite{[16]} accompanied by a 
striped-SDW long range order at $T_{SDW}$.  In fact, in general,
$T_{s}>T_{SDW}$,~\cite{[16]} supporting such a view.  This could also lead 
to a nematic instability, a possibility that has already been
proposed in the FePn context.~\cite{[17]}

(iii) Finally, these dynamical {\it short-ranged} spin correlations should
 survive the doping-induced melting of the SSDW.  These nearest- and diagonal
neighbor correlations are attractive candidates for the ``glue'' that leads to 
short-coherence length U-SC in FePn.  In fact, the U-SC gap function should 
then have neartest neighbor (n.n) and diagonal components: these {\it exactly}
correspond to the ex-$s$ and $s_{xy}$-pair components. Interestingly, rigorous
analysis {\it requires} that {\it both} co-exist in FePn.~\cite{[18]} This is
precisely the in-plane, {\it nodeless} part of the SC gap function that we have 
proposed in Ref.~\onlinecite{[19]} as an instability of the incoherent 
``normal'' state in doped FePn.  Since the pair wavefunction has only n.n 
and next-n.n components as above, this also affords a way to rationalize 
the short SC coherence length, high upper critical fields, and Uemura 
scaling, all of which are observed in FePn.~\cite{[20]}  

Hence, in our work, the anomalous spin susceptibility arises from effectively 
localized, short-ranged and sizably frustrated spin correlations above
$T_{SDW}$ in the 1111-FePn.  That the FePn are bad metals (the drop in
$\rho(T)$ at low $T$ in some of the FePn, especially in $LaO_{1-x}FeAsF_{x}$, 
actually {\it correlates} with the opening of the spin gap,~\cite{[fuchs]} 
and, cannot be taken as evidence of FL behavior)~\cite{[6]} means that these 
short-ranged spin correlations are further damped by the dynamical charge 
fluctuations in reality. As a further check of our analysis, we notice that, 
choosing the spin gap, $\Delta\simeq 150-200$~K, from our NMR result, the 
$dc$ resistivity is expressed as 
$\rho_{dc}(T)\simeq \rho_{0}+CT$e$^{-\Delta/k_{B}T}$, in good agreement with
the observations for $x=0.1$ in $LaOFeAs$;~\cite{[fuchs]} interestingly,
extraction of $\Delta$ from the dc resistivity gives $\Delta=164$~K for
$x=0.1$, in good agreement with our estimate from NMR.  When the spin gap 
closes, the dc resistivity recovers its characteristic linear-in-$T$ behavior,
as indeed seen.~\cite{[fuchs]} This is again
reminiscent of the evolution of $\rho_{dc}(T)$ in underdoped cuprates with
increasing hole doping.~\cite{[4]}   
In a ``Mottness'' scenario, with relatively stable local moments 
(this should not be confused with small sublattice {\it magnetization} in 
FePn, which can be drastically lowered by frustration),~\cite{[2]} the 
{\it effective} Heisenberg limit should be qualitatively correct below an 
exchange scale renormalized {\it downward} by metallicity.  We emphasize that 
this argument does not constitute a strict mapping from our DMFT results to 
an effective frustrated Kondo-Heisenberg model (for a more formal mapping, 
see Ref.~\onlinecite{[2]}); however, the latter does allow deeper insight 
into the DMFT results.

Finally, at lower $T$ for $x\simeq 0.1$, the dimensional crossover from 
$D=2$ to $D=3$ should occur.  This is what leads to the deviation of the 
low $T$ susceptibility from the linear-in-$T$ behavior.  Our finding of 
an approximate $\chi(T)\simeq T^{1.8}$ law in this regime (in nice agreement 
with the $x=0.1$ La-based 1111-FePn) suggests multi-orbital effects at work: 
in particular, we argue that small inter-layer couplings (hopping between 
$d_{z^{2}}$ orbitals on neighboring layers, and concomitant superexchange,
$J_{\perp}$), setting in at low temperatures, lead to anisotropic $D=3$ 
physics, giving an approximate $\chi(T)\simeq T^{2}$ law, smoothly evolving 
from the linear-in-$T$ variation at higher $T$.
  
Our analysis thus leads to a very different picture compared to the one
obtained from an itinerant view.~\cite{[8]}  In the itinerant view, spin 
gap generation requires rather involved extensions of the HF-RPA approach 
hitherto used therein,~\cite{[21]} and these should bring the theoretical 
description closer to the limit considered here. Also, evolution of 
$\chi(T)\simeq T$ to the $T^{1.8}$ dependence with doping (for $x=0.1$ 
and beyond see Ref.~\onlinecite{[6]}) remains to be clarified there.  In our 
``strong coupling'' picture, this behavior arises from the $2D-3D$ crossover
which should set in with increasing $x$ at low $T$ in the quantum paramagnetic
metal. Using LDA+DMFT, we show how {\it both}, the $\chi(T)\simeq T$ 
at high $T$, as well as its $T^{2}$-like behavior for low $T$ in the $x>0.06$
regime, can be obtained in a strongly correlated picture.  Both the 
linear-in-$T$susceptibility {\it and} the spin gap are thus associated with 
the wide ranged (in $T$) fluctuational contributions associated with 
short-ranged, geometrically frustrated spin correlations in the quantum 
disordered regime of an {\it effective}, doped $J_{1}-J_{2}$ antiferromagnet. 
Our view is also fully consistent with ARPES,~\cite{[22]} optical~\cite{[23]} 
and $\mu$SR~\cite{[24]} data, all of which are consistent with a strongly 
correlated metal giving way to U-SC.        
            
In conclusion, we have studied the temperature dependence of the 
uniform spin susceptibility, $\chi(T)$, for the 1111-FePn using the 
first-principles LDA+DMFT method.  Good semiquantitative agreement with 
published experimental data testifies to the relevance of short-ranged and 
dynamical multi-band electronic correlations in FePn.  The same approach 
also yields a satisfying description of the NMR $[T_{1}T]^{-1}$ as a function 
of $T$ in an earlier work.~\cite{[19]}  We thus identify {\it both}, 
the spin gap in NMR {\it and} $T$ dependence of $\chi(T)$, as 
manifestations of the wide (in $T$) fluctuational regime of a sizably 
frustrated magnet between its actual ordering scale ($T_{SDW}$) and the 
mean-field crossover ($T_{MF}$).  Our study pinpoints the crucial roles of 
strong multi-orbital electronic correlations and sizable geometric frustration
to understand magnetic fluctuations in the incoherent metallic state of the 
1111-Iron Pnictides.

\acknowledgments
The Authors thank the MPIPKS, Dresden, for hospitality and financial
support.


\begin{thebibliography}{19}

\bibitem{[1]} 
I.I. Mazin and J. Schmalian, arXiv:0901.4790.

\bibitem{[2]} Q. Si and E. Abrahams, Phys. Rev. Lett. {\bf 101},
076401 (2008); ibid. J. Wu, P. Phillips, and A.H. Castro Neto, 
Phys. Rev. Lett. {\bf 101}, 126401 (2008); G. Baskaran, J. Phys. 
Soc. Jpn. {\bf 77}, 113713 (2008); Q. Si, E. Abrahams, J. Dai, 
and J.-X. Zhu, arXiv:0901.4112.

\bibitem{[3]}  C. Xu, M. M\"uller, and S. Sachdev,
Phys. Rev. B {\bf 78}, 020501(R) (2008).

\bibitem{[4]} H. He {\it et al.}, 
Phys. Rev. Lett. {\bf 86}, 1610 (2001); ibid. J. Bobroff {\it et al.}, 
Phys. Rev. Lett. {\bf 78}, 3757 (1997).

\bibitem{[5]} Y. Nakai, S. Kitagawa, K. Ishida, Y. Kamihara, M. Hirano, 
and H. Hosono, arXiv: 0810.3569.

\bibitem{[6]} G.-M. Zhang, Y.-H. Su, Z.-Y. Lu, Z.-Y. Weng, D.-H. Lee, 
and T. Xiang, arXiv:0809.3874.

\bibitem{[kling]} R. Klingeler, N. Leps, I. Hellmann, A. Popa, C. Hess, 
A. Kondrat, J. Hamann-Borrero, G. Behr, V. Kataev, and B. B\"uchner, 
arXiv:0808.0708.

\bibitem{[7]} M.L. Foo, Y. Wang, S. Watauchi, H.W. Zandbergen, T. He, 
R.J. Cava, and N.P. Ong, Phys. Rev. Lett. {\bf 92}, 247001.

\bibitem{[8]} M.M. Korshunov, I. Eremin, D.V. Efremov, D.L. Maslov, 
and A.V. Chubukov, arXiv:0901.0238.

\bibitem{[9]} A.I. Coldea {\it et al.}, 
Phys. Rev. Lett. {\bf 101}, 216402 (2008).

\bibitem{[10]} G. Kotliar, S.Y. Savrasov, K. Haule, V.S. Oudovenko, 
O. Parcollet, and C. A. Marianetti, Rev. Mod. Phys. {\bf 78}, 865 (2006).

\bibitem{[11]} K. Haule, J. H. Shim, and G. Kotliar, Phys. Rev. Lett.
{\bf 100}, 226402 (2008).

\bibitem{[12]} L. Craco, M. S. Laad, S. Leoni, and H. Rosner,
Phys. Rev. B {\bf 78}, 134511 (2008); ibid
M.S. Laad, L. Craco, S. Leoni, and H. Rosner,
Phys. Rev. B. {\bf 79}, 024515 (2009).

\bibitem{[13]} L. Craco and M.S. Laad, arXiv:0903.1568.

\bibitem{[19]} M.S. Laad and L. Craco, arXiv:0902.3400.

\bibitem{[rmpg]} A. Georges, G. Kotliar, W. Krauth and M. Rozenberg, Revs. Mod. Phys. {\bf 68}, 13 (1996).

\bibitem{[14]} U. Chatterjee {\it et al.}, 
Phys. Rev. B {\bf 75}, 172504 (2007).

\bibitem{[15]} D. Hinzke, U. Nowak, and D.A. Garanin, 
Eur. Phys. J. B {\bf 16}, 435 (2000).

\bibitem{[16]} C. de la Cruz {\it et al.}, 
Nature {\bf 453}, 899 (2008). 

\bibitem{[17]} S.A. Kivelson and H. Yao,
Nature Materials {\bf 7}, 927 (2008).

\bibitem{[18]} W.-L. You, S.-J. Gu, G.-S. Tian, and H.-Q. Lin,
arXiv:0807.1493.

\bibitem{[20]} Y.J. Uemura, arXiv:0811.1546.

\bibitem{[fuchs]} G. Fuchs {\it et al.}, 
arXiv:0902.3498.  

\bibitem{[21]} P. Monthoux and G.G. Lonzarich,
Phys. Rev. B {\bf 66}, 224504 (2002).

\bibitem{[22]} L. Wray {\it et al.}, 
Phys. Rev. B {\bf 78}, 184508 (2008).

\bibitem{[23]} A.V. Boris {\it et al.}, 
Phys. Rev. Lett. {\bf 102}, 027001 (2009).

\bibitem{[24]} R. Prozorov {\it et al.}, 
arXiv:0901.3698. 


\end{thebibliography}
\end{document}